# Teaching Creativity Using a Realistic Multi-User Operation: Packet Tracer

**Muhammad Khairul Ezad Bin Sulaiman**

Jabatan Teknologi Maklumat dan Komunikasi
Politeknik Tuanku Syed Sirajuddin
Pauh Putra, 02600 Arau, Perlis
khairul_ezad@yahoo.co.uk

**Abstract**

Multi-user capabilities in Cisco Packet Tracer provide an incentive for immersive realistic learning to increase the efficiency of teaching and learning in the networking class. It is difficult to evaluate the configuration skills of the students in the classroom, particularly during laboratory class, while teaching networking course, the most problems for lecturers. The multi-user functionality that uses the one-to-many remote relay server concept allows lecturers to remotely control and test many students at one time. The key purpose of this paper is to incorporate creativity in teaching by using realistic multi-user practises in the networking class. In this study, the multiuser operation was developed and used via an existing LAN link in the classroom. The focus of the operation is the simple configuration of network equipment such as routers and switches. This practise was used during the laboratory session, encouraging lecturers to establish relationships based on a large audience. The professor is able to evaluate the success of the students during the execution of the activity and to ensure that each student engages in the activity. As a result of introducing this approach, the curiosity of students in technical subjects can be improved and student success can be conveniently tracked on the part of the lecturer. Lecturers are able to monitor the configuration capabilities of students closely to ensure that each student engages in the laboratory class. The lecturer will train students not only to develop their configuration and troubleshooting abilities by using this teaching form, but also to accelerate the pace at which students need to fix network problems.

## 1. Introduction

In order to empower all students with awareness, expertise and mindset to face the global demands and requirements of the ICT industry today, there are different teaching approaches used to conduct in the class. The bulk of the networking course at the Polytechnic in Malaysia is intended to provide students with perceptual, psychomotor and productive learning results based on the taxonomy of the Bloom. This day's advancement in Information Networking Technologies brings so much complexity to educators. Therefore, educators need to build creative curricula and teaching resources that can help students grasp the nuances of information and communication technologies. It has also allowed innovative and more effective methods to be created for teaching environments. Owing to today's students unable to keep their focus for too long, the lecture techniques that most teachers used to perform in the class no longer function. . Traditional lecture-style teaching environment which the teacher had the central role in educational activities and student were merely listeners is unable to attract students to give full attention in the class anymore (Musheer, Sotnikov and Heydari, 2012). A study finds that undergraduate students in classes with traditional stand and deliver lectures are 1.5 times more likely to fail than students in classes that use more stimulating (Bajak, 2014).

The teaching methods that encourage students to be actively engaged in teaching and learning processes are more appropriate nowadays. The teaching approaches that turned students into active participants rather than passive listeners reduced failure rates and boosted scores on exams by almost one-half a standard deviation, according to Bajak (2014). There is, however, no rule book about which instructional strategies better complement what skills or material is being learned. It depends on the students at that time and the predicted benefit from the learning result. In addition, the anticipated learning result for the networking courses is that students can understand the principles of networking theory and can conduct hands-on realistic lessons. The curriculum of networking courses at Polytechnic has merged with the curriculum of Cisco Networking Academy technical qualifications such as CCENT and CCNA so as to provide the opportunity for the students to sit for professional qualification exams (Curriculum development division, 2010). This will bring benefit to the students and guarantee that the expertise and skills learned during this curriculum are directly applicable to the needs of the ICT industry in networking. This programme is to train students in today's marketplace for a competitive advantage.

In general, all networking course syllabuses are meant to teach students with theoretical knowledge and practical ability. Students will learn theoretical knowledge in the classroom through the lesson, while students will learn in the experimental classroom for the practical portion. Typically, students find networking topics technical and dull since they consist of a complicated and changing subject (Zhang, Liang and Ma, 2012). Students who research networking must be able to clarify the basics of how computer networks operate, according to Airi and Anderson (2017). Students and lecturers continue to devote several hours engaged with functional applications planning, configuring and integrating a virtual network in order to produce these learning effects. In addition, to test the comprehension of students and to ensure that students meet the learning objectives of the course, lecturers need to perform the examination according to the appraisal tasks defined in the syllabus. Basically, training can be performed to test their understanding of theories and functional abilities. For the practical part, though, it is difficult for the professor to assess the students' interpretation and to evaluate it. To ensure that all students are able to correctly execute a setup on network equipment is very necessary for the lecturer. From the previous observation, during the conduct of the networking class, the difficulties of the professor are that students are not completely interested in the networking class, it is difficult to determine the student development of functional skills and students use a long period to finish the simple setup.

## 1.1 Functionality for Multiusers

One of the features of Packet Tracer that enables multiple point-to-point (peer) links between multiple Packet Tracer instances is the multi-user communication. For students to study in roups, lecturers are able to establish a set of activities that will promote greater social contact between students. The multi-user role facilitates thrilling interactive and competitive experiences, offers the ability to advance from person to social learning and provides opportunities for communication, competition, remote interactions between teachers and students, social networking and gaming (Packet Tracer Data Sheet, 2010). This functionality enables users to connect to a peer cloud using a quick drag and drop cloud icon. Each multiuser cloud supports configurations of one-to-one, many-to-one and many-to-many peer connections. Two sides, the server or client architecture, were interested in the multiuser operation log. The key file may be hosted on a server or lecturer PC, and each student uses the client or student side file to connect to the primary Packet Tracer file hosted on the lecturer PC. The multi-user capabilities allow up to 60-75 users to link clients to a single operation concurrently over the same wired LAN network (Musheer, Sotnikov and Shah Heydari, 2011).

## 1.2 Configuring Network Equipment

Computers, cell phones, peripherals, and even IoT devices are linked by a network. The key networking essentials are switches, routers, and wireless access points. Via them, network-connected communication devices can interact with each other and with other networks, such as the Internet (Connect employees and offices, 2017). Therefore, the basic configuration is required for all network devices, especially switches and routers, to support network communication. Basic router or transfer network systems provide host name settings for authentication, security passwords, and the assigning of IP addresses to connectivity interfaces (Cisco Systems, 2008). It is an initial router and switch setup that students can study for the syllabus in networking class, such as in the course Switching and Routing Basics. It is necessary for students to understand the theory principles and to be able to configure the network connectivity as an initial to set up.

# 2. Materials And Methods

Switching and Routing Essentials was the path selected for the introduction of multi-user operations. This course is made up of 30 students in semester 4. The syllabus contained the initial router and switch setup to be mastered by the students in Switching and Routing Basics. Because of that, this course and students were selected in this study as a respondent. The lecturer's observation and pre-post questionnaire was used to provide students with a perspective of their knowledge using realistic multi-user exercises to understand the configuration of simple network equipment. During the preceding class, the questionnaire adapted from (Smith, 2011) and (Šimandl, 2015) is based on the issue indication. The questionnaire is intended to test the comprehension of students in the design of simple network equipment for the definition and configuration of theory, the extent of trust of students to complete the activities, the efficiency of students to complete each task as well as their involvement in the activities.

Students learnt about the core network interface design for theory definition and configuration prior to the introduction of the multi-user functional tasks. The theory principle taught during the two hours theory class and the configuration is performed according to the lesson timetable allotted during 2 hours practical class. Using the standard Packet Tracer and even using the actual computers, the setup process was carried out. After the lab exercise was completed, the pre-test questionnaire was immediately administered to the students to determine the comprehension of the students after each activity. The bulk of the time was used by each student, with an average of thirty minutes left to allow the students to complete the questionnaire. In this analysis, the programme used to create the operation file is Packet Tracer version 7.0.0. It is a creative network setup, a modelling platform that allows students and lecturers to develop their desktop or mobile computer networking configuration skills. Packet Tracer is an important learning method for Cisco Networking Academy courses that enables students to experiment with network activity for activities and evaluation (Packet Tracer Data Sheet, 2013). In addition, this simulation programme can be used in the class due to the curriculum of networking courses in Polytechnic combined with the Cisco Networking Academy curriculum and the lecturers who are certified as Cisco instructors are legal. Archana (2015) notes that Packet Tracer offers capabilities for modelling, visualisation, authoring, evaluation, and collaboration and promotes the teaching and learning of dynamic concepts of technology. Simulation software is a special application that helps students without the need for actual network equipment to build and customise a network (Makasiranondh et al., 2010. While simulators do not provide students with substantial functional skills, such as cabling and physical networking, they are a valuable, cost-effective addition to teaching programmes (Šimandl and Vaníček, 2015). Al-Holou et al. (2000), however, contend that this programme may be used to increase the comprehension of a student as it allows the simulation of limit conditions that introduce dangers in real-life situations but contribute to a deeper understanding of concepts. Simulation programme allows for student-centered, inquiry-based teaching and learning, according to Rutten (2014). It is not only learned personally or in small groups as necessary for studying, but also at the entire level of the curriculum. Multi-user capabilities of Packet Tracer have been used for this analysis. The activity is a multi-user activity that was planned to allow 20 to 30 minutes for students to complete it. The first task is to test the students' capacity to customise the simple configuration. The task assumes the students in the previous class have mastered the basic setup.

## 2.1 Multiuser Environment Setup

Two client-server network topology environments have been developed. Server network topology is for lecturer side use to perform the students. Since the topology of the client network is for the student side. By creating connections to the lecturer side of the operation, the students may launch a multiuser Packet Tracer activity on their respective desktop or laptops. At some point, the faculty side should be able to determine the setup of the student and save the progress of the student.

The lecturer-side network topology was generated by connecting routers in the Device-Type Selection Box to the Multiuser Link. Pick the Remote Network Cloud and build it in the workspace of the Packet Tracer. This cloud provides a single point of entry for the student. The number of remote cloud networks depends on the number of students in the current class. Many multi-user entry points are approved for the same user. After the setup was completed, one computer connected to the router to run connectivity checks between the lecturer side and the student side. Network topology created by linking a router to the Remote Network cloud on the student side. The Remote Network cloud is an entry point for linking the student side to the faculty side. Linked to the end devices is another side of the router. For this evaluation, one laptop and a server are the final machines used. The cloud of the Remote Network is also known as a peer link. The first peer is called by default as Peer0 in the network topology.

Both teacher and student computers must be linked to the same local area network to create a local multi-user connection (LAN). Using the Extensions menu in the Packet Tracer to customise multi-user settings for lecturers and pupils. By matching IP address, port, cloud name and password parameters between lecturer and student, the relation between lecturer and student is achieved. For any peer link on the part of the lecturer, an incoming connection must be ready. Before the student can construct a connection to the lecturer, the student side must have an IP address, port number and password used by the lecturer side.

The pupil, instructor-side IP address, port number, and password must be given by the lecturer. To create a connection, the student side requires these three specifics. A multi-user connection may be established on any port, as suggested by Smith and Bluck (2010), with ports 38000 to 38999 presented by default. The port number is set by default in this report. For the student-side Remote Network peer, the link type must be changed to Outgoing.

The Peer Cloud on the student side and the cloud on the host side will turn blue to check the connectivity. The linking light between the network devices and the cloud will change from amber to green after a short time. The multi-user link, meanwhile, is now developed and ready for testing.

**2.2 Practical Class Operation**

For all the students' connections, a single activity file was used and a single lecture file was used that would accept all the connections. The simple network interface setup was used as a first exercise to introduce students to a realistic learning approach for multiple users. The multi-user student's file was provided simultaneously to all 15 students. The lecturer explained the comprehensive guidance about how students should relate to the lecturer's file and the setup tasks they need to complete. Students all together must connect to the lecturer file to begin the communication. To ensure that all students fulfil this role through all the peer cloud in the lecturer file all transform to blue colour, the lecturer is able to check.

There are three activities that students have to finish before students link to the cloud. The basic setup is to execute the first functions. The simple setup on the router that connects to the cloud needs to be configured for studies. The basic configuration was:

  i. Hostname R0
 ii. Secret password cisco
iii. Line console password cisco
 iv. Line vty password cisco
  v. Disable DNS lookup
 vi. Configure message-of-day banner-RESTRICTED ZONE

The second role is to make students customise the IP address on the router interfaces. Serial Interface and LAN Interface were used in the protocols. The tutor provides the IP addresses. Students have to set the IP address of the desktop and server end computers for the last activities. The IP addresses are given by the teacher as well.

The instructor will execute the ping from PC INSTRUCTOR to the end devices on the LAN student side to check the connectivity. The outcome ought to be effective. In order to check the configuration of each student, the teacher can access the configuration of each student via Telnet on each student's router.

## 3. Result and Discussion

Once the multi-user practise has been applied to the students in the class, it is important to address the question of involving students in the class. The multi-user operation will encourage lecturers and students to learn in a more interactive way and facilitate real-time interaction between lecturers and students. A fast, engaging and extensible experience that can be used to facilitate student involvement in lectures will fill a void by providing multi-user activity as part of a networking course. Students can conveniently link from the observation to the lecturer-side network. The lecturer will ensure the turn to blue colour of all the peer cloud on the lecturer side. From this, the professor will promise that all students are interested in the practical class. It may attract learners to participate in the learning phase in the class through the use of this exercise. The multi-user exercise will facilitate student involvement in the classroom and at the same time detect if any students are attempting to avoid engaging in the classroom activity. This practise also means that the realistic activity includes all pupils. This will eliminate the difficulties that other learners have copied instead of doing by themselves as some have learned. Configuring the core framework is the first task of the multi-user operation. Configure basic network system parameters, such as router and switch, including global parameter configurations, protocols, and access to commands. Students need to instal the initial settings on a router or switch until it can be used for network communication. After introducing the multi-user practise, evaluation by lecturers was carried out to test the impressions of students and questionnaires were given to students at the end of the class. The questionnaires are based on four (4) factors that ask about the students' comprehension, their faith in configuration, students' effectiveness and student involvement in the curriculum. Students were asked to give answers to the questions by offering yes or no answers. Table 1 displays the input findings from semester 4 of 15 students in class.

Table 1. Students' perception after implementation of Multiuser activity

| CATEGORY | QUESTIONS | YES (%) | NO (%) |
| --- | --- | --- | --- |
| Understanding | I properly understand the concepts of basic network configuration. | 93.3 | 6.7 |
| Confidence | I can configure the network devices on my own. | 86.7 | 13.3 |
| Productivity | I can finish the configuration tasks on time. | 73.3 | 26.7 |
| Engagement | Multiuser actively encouraged me to fully engage in the configuration laboratory exercise. | 100.0 | 0 |

Table 2: Students' perception before implementation of Multiuser activity

| CATEGORY | QUESTIONS | YES (%) | NO (%) |
| --- | --- | --- | --- |
| Understanding | I properly understand the concepts of basic network configuration. | 60.0 | 40.0 |
| Confidence | I can configure the network devices on my own. | 66.7 | 33.3 |
| Productivity | I can finish the configuration tasks on time. | 53.3 | 46.7 |
| Engagement | Multiuser actively encouraged me to fully engage in the configuration laboratory exercise. | 66.7 | 33.3 |

The result shows that the percentage of comprehension, trust level, efficiency and interaction of all facets after multi-user operation execution rises comparably with before execution. That means this practise has led to the awareness of the network configuration of students and the direct development of their functional skills in the network. In addition, students can also be inspired to learn of ways to solve problems effectively. From the observation, because of not remembering the configuration command lines, typically students are unable to complete the laboratory operation during laboratory class hours, others are the sloppy error in configuration commands, and the worst is that students do not read the instruction in the laboratory sheet issued. For these purposes, it would take more time for students to complete their assignments. The tutor is able to track the learners automatically during the multi-user process. In order to complete the assigned assignments, it will implicitly require students to act efficiently and accurately to solve the problems the fastest.

Students collected input and noticed that students were genuinely interested and had not attempted to prevent difficulties by observation. Students also indicated that they were interested in using the multiuser approach to perform realistic tasks. For multi-user operation, the setup can be appreciated when they were implicitly required to do themselves correctly. Thus, they would check by themselves whether they had worked and hadn't. They became autonomous and their knowledge of the dilemma had improved. Lecturers are able to specifically and automatically track the pupils. The direct reaction of the lecturer to recognise the student's failure and error will increase the consistency of the teaching and learning process.

## 4.0 Conclusion

At the Diploma level of Politeknik, this paper outlined several difficulties involved in teaching networking courses. Therefore, lecturers can train students not only to develop their configuration and troubleshooting skills by using this process, but also to increase configuration speed. This practise also means that the realistic activity includes all pupils. This will eliminate the difficulties that other learners have copied instead of doing by themselves as some have learned. The multi-user characteristics that may help the lecturer to determine the student's unique network will assist the lecturer to recognise certain students with a challenge and lack of configuration. From this, the lecturer will give the students direct feedback and response. The multiuser approach helps the professor to see how much student's setup and troubleshooting abilities have advanced. From this, the lecturer will give the students direct feedback and response. From the outcome, it can be inferred that the multi-user operation is ideal as another teaching and learning tool to be introduced in the class. Lessons were judged by the students themselves as useful and appealing, so there was not much time to be bored.

Lecturer introduces students to the daunting challenge in future work instead of just presenting students with do as instructed activities that provide the answers in depth, then students will complete the exercise step by step and they will be bored. More creative ideas could also be sought by the professor, and students became more involved in this course. Multi-user activity aimed to inspire students to accomplish the challenge by themselves during a laboratory class as a practical exercise to enable students to include their ingenuity with a solution.

# REFERENCES

Airi, P., & Anderson, P. K. (2017). Cisco Packet Tracer as a teaching and learning tool for computer networks in DWU. *Contemporary PNG Studies: DWU Research Journal Vol. 26 May 2017*

Al-Holou, N., Booth, K. K., & Yaprak, E. (2000). Using a computer network simulation tools as supplements to the computer network curriculum. *In Frontiers in Education Conference, 2000. FIE 2000. 30th Annual (Vol. 2, pp. S2C-13). IEEE.*

Archana, C. (2015). Analysis Of Ripv2, Ospf, Eigrp Configuration On Router Using Cisco Packet Tracer. *International Journal Of Engineering Science And Innovative Technology (Ijesit) Volume, 4.*

Bajak, A. (2014). Lectures aren't just boring, they're ineffective, too, study finds. *Science Insider.*

Cisco Systems, Inc., (2008). Cisco Secure Router 520 Series Software Configuration Guide. *Chapter 1 : Basic router configuration.*
Retrieved October 23, 2016, from https://www.cisco.com/c/en/us/td/docs/routers/access/500/520/software/configuration/guide/520_SCG_Book.pdf

Connect employees and offices. (2017). Networking basics: what you need to know. Retrieved October 23, 2016, from https://www.cisco.com/c/en/us/solutions/small-business/resource-center/connect-employees-offices/networking-basics.html

Curriculum development division (2010). Diploma In Information Technology (Networking). *Department of polytechnic education, Ministry of higher education in Malaysia*

Makasiranondh, W., Maj, S. P., & Veal, D. (2010). Pedagogical evaluation of simulation tool usage in Network Technology Education. *World transactions on engineering and technology education,* 8 (3), 321-326.

Maor, D., & Fraser, B. J. (2005). An online questionnaire for evaluating students' and teachers' perceptions of constructivist multimedia learning environments. *Research in Science Education*, 35 (2), 221-244.

Musheer, A., Sotnikov, O., & SHAH HEYDARI, S. (2011). Packet Tracer as an Educational Serious Gaming Platform. *In Comunicação apresentada no (a) Seventh International Conference on Networking and Services* (Vol. 22, pp. 299-305)

Musheer, A., Sotnikov, O. and Heydari, S. S. (2012). Multiuser Simulation-Based Virtual Environment for Teaching Computer Networking Concepts. *International Journal on Advances in Intelligent Systems,* Vol 5 no 1 & 2, year 2012.

Packet Tracer Data Sheet. (2010). Cisco Packet Tracer data sheet. CISCO Networking Academy. Retrieved October 21, 2016, from http://www.cisco.com/c/dam/en_us/training-events/netacad/course_catalog/docs/Cisco_PacketTracer_DS.pdf

Packet Tracer Data Sheet. (2013). Cisco Packet Tracer data sheet. CISCO Networking Academy. Retrieved October 23, 2016, from http://www.cisco.com/c/dam/en_us/training-events/netacad/course_catalog/docs/Cisco_PacketTracer_DS.pdf

Rutten, N. P. G. (2014). Teaching with simulations (No. 14-317). *Universiteit Twente*.

Šimandl, V., & Vaníček, J. (2015). The use of inquiry based education in a simulated software environment in pre-service ICT teacher training. *International Journal of Information and Communication Technologies in Education,* 4 (1), 5-15.

Smith, A., & Bluck, C. (2010). Multiuser collaborative practical learning using packet tracer.

Smith, A. (2011). Classroom based Multi-player Network Simulation. *In the seventh international conference on Networking and Services ICNS*.

Zhang, Y., Liang, R., & Ma, H. (2012). Teaching innovation in computer network course for undergraduate students with packet tracer. *IERI Procedia*, *2*, 504-510.